# Conservation Laws and Spin System Modeling through Principal Component Analysis

David Yevick
Department of Physics
University of Waterloo
Waterloo, ON N2L 3G7

**Abstract:** This paper examines several applications of principal component analysis (PCA) to physical systems. The first of these demonstrates that the principal components in a basis of appropriate system variables can be employed to identify physically conserved quantities. That is, if the general form of a physical symmetry law is known, the PCA can identify an algebraic expression for the symmetry from the observed system trajectories. Secondly, the eigenvalue spectrum of the principal component spectrum for homogeneous periodic spin systems is found to reflect the geometric shape of the boundary. Finally, the PCA is employed to generate synthetic spin realizations with probability distributions in energy-magnetization space that closely resemble that of the input realizations although statistical quantities are inaccurately reproduced.

**Introduction:** Machine learning can be employed to isolate, predict or simulate features of data in physical systems. Consequently, it can be applied to broad classes of problems although at the cost of introducing numerical uncertainties typically associated with stochastic optimization problems. [1][2][3] Despite this limitation, which is further augmented by the presence of metaparameters with problem dependent optimal values, machine learning has been successfully applied to many areas of condensed matter physics. The first of these employed PCA [4][5][6][7] and neural network [8][9][8][10][11][12][13][14] techniques, which were typically employed to identify different phases of Ising and related spin models. Subsequent research employed both supervised learning, which requires training the algorithm with preexisting sets of input and output data pairs [15][16][17][18][12][19][20] and unsupervised learning in which the training phase is absent [5,7,21–27]. The latter techniques are typified by cluster analysis, which classifies data into groups according to perceived similarities, and feature extraction, which projects the data set onto a low-dimensional space while still preserving essential characteristics of the original data. Examples of specific problems that have been examined with machine learning procedures are the simulation of quantum systems [18,28–32], the prediction of crystal structures [33,34], the approximation of density functionals [35] and the solution of quantum impurity problems. [36] Additionally, machine learning can be employed to identify system properties and in particular can identify manifest and hidden order parameters and different phases or states of a system, especially when supplemented by additional knowledge of system properties such as locality, translational symmetry and symmetry breaking. [5,6,14,20,25,37,38]

A recent application area of machine learning is the identification of underlying symmetries or laws governing the evolution of a physical system. Such analyses extend the functionality of machine learning beyond standard computer vision or language processing implementations in that they enable not only the prediction or reproduction but also the interpretation of observed phenomena. For example, the underlying physical laws or symmetries that describe the behavior of a physical system can in certain cases by identified and even formulated algebraically.[9][39,40]   This indicates that machine learning can potentially detect the underlying structure of a system, which could potentially be employed to justify or to formulate novel physical concepts.

This paper implements unsupervised learning through the PCA [4], which leverages singular value decomposition to identify orthogonal linear combinations of the data variables with the property that the data exhibits the largest variance when projected onto the lowest order principal component and successively smaller variances along successive components.  Applying cluster analysis to the distribution of data projected along the lowest order principal components then enables the identification of phase transitions in Ising and other spin systems. [4][5][6][7]   Here, however we analyze three rather different aspects of the PCA representation of physical systems.   In the first of these, we demonstrate that the method can be employed to detect physical symmetries if these can be expressed in simple algebraic form.  Secondly, we find that in periodic lattices the magnitudes of the non-dominant PCA eigenvalues are influenced by the geometric shape of the computational grid and are effectively artifacts of the distortion of the grid upon mapping onto a sphere.  Finally, we demonstrate that if a PCA is trained with data from a stochastic lattice model, its outputs can be easily manipulated to produce synthetic data that, unlike some other machine learning procedures, preserves the joint distribution of the data in energy and magnetization.  However, this is accompanied by a substantial inaccuracy in the prediction of statistical quantities.

**Implementation:**  To introduce and clarify the methods discussed in the results section in the most concrete fashion, the core elements of the associated MATLAB$^©$ programs are first presented.  In the first PCA application, following the general prescription of [39],  a large number of two-dimensional harmonic oscillator trajectories are generated. The positions at each recorded point along these trajectories are stored in rows of the matrix **particlePositionSaveRC** while the associated velocities are similarly placed in **particleVelocitySaveRC**.  In MATLAB$^©$ with the syntax conventions of [41][42] this is simply performed with the following code lines (the code is parallelized by substituting **parfor** in place of **for**).

```
numberOfRealizations = 500;
timeSpanR = [0 : 4000];
particlePositionSaveRC = [ , ];
particleVelocitySaveRC = [ , ];

for realizationLoop = 1 : numberOfRealizations
   angle = rand * 2 * pi;
   particlePositionRC =  [ cos( angle ), -sin( angle ); ...
   sin( angle ), cos( angle ) ] * [ rand + 2; 0 ];
   particleVelocityRC = ( 2 * rand( 2, 1 ) - 1 ) / 4;
   initialConditionsRC = [ particlePositionRC; particleVelocityRC ];
   [ timeVectorC, outputVectorRC ] = ode45( @oscillatorFunction, timeSpan, initialConditions );
   particlePositionSaveRC = [ particlePositionSaveRC; outputVectorRC(:, 1:2) ];
   particleVelocitySaveRC = [ particleVelocitySaveRC; outputVectorRC(:, 3:4) ];
end
```

where **oscillatorFunction( )** is defined as

**function outputC = oscillatorFunction( timeR, xR )**

```
   outputC = [ xR(3);  xR(4);  –xR(1);  –xR(2) ]
```

Note that the trajectories, while random in direction and velocity, are launched in a region displaced from the position origin.  Once the positions and velocities are acquired, a data matrix consisting of the 14 values  $x^2, y^2, v_x^2, v_y^2. xv_y, yv_x, xy, xv_x, yv_y, v_xv_y, x, y, v_x, v_y$  at each point of the trajectories is constructed as follows

```
dataRC = [ particlePositionSaveRC(:, 1).^2, particlePositionSaveRC(:, 2).^2, …
   particleVelocitySaveRC(:, 1).^2, particleVelocitySaveRC(:, 2).^2, …
   particlePositionSaveRC(:, 1) .* particleVelocitySaveRC(:, 2),…
   particlePositionSaveRC(:, 2) .* particleVelocitySaveRC(:, 1), …
   particlePositionSaveRC(:, 1) .* particlePositionSaveRC(:, 2), …
   particlePositionSaveRC(:, 1) .* particleVelocitySaveRC(:, 1), …
   particlePositionSaveRC(:, 2) .* particleVelocitySaveRC(:, 2), …
   particleVelocitySaveRC(:, 1) .* particleVelocitySaveRC(:, 2), …
   particlePositionSaveRC(:, 1), particlePositionSaveRC(:, 2), …
   particleVelocitySaveRC(:, 1), particleVelocitySaveRC(:, 2) ];
```

The principal components are then generated for this data set by a call to the MATLAB **pca( )** function which returns and displays the principal components as the columns of the matrix **coeffRC**, with the corresponding eigenvalues (variances) stored in the column vector **latent**

**[ coeffRC, scoreRC, latentC ] = pca( dataRC )**

The remaining two applications of the PCA relate to the Ising model.  A data set is constructed at a temperature $T$ by populating consecutive rows of a single large $M \times N^2$ matrix, **realizationSave** with $M$ consecutive Wolff samples of a $N \times N$ Ising system, where the successive spin matrices are flattened into vectors of values 1 or -1 corresponding to up and down spins, respectively.  The PCA can then be implemented after rescaling this data matrix so that each column possesses a zero mean (in the example below the means of the columns are typically $< 0.01$, so that omitting this normally required step affects the computed specific heat only in the third decimal place).  In this example the PCA is instead implemented though the MATLAB single value decomposition routine as

```
realizationSaveMeanR = mean( realizationSaveRC );
rescaledRealizationRC = realizationSaveRC - realizationSaveMeanR;
[ pcaScoreRC, pcaLatentRC, pcaCoeffRC] = svd( rescaledRealizationRC, 'econ' );
```

The slightly redefined principal component vectors are stored as the columns of the matrix **pcaCoeffRC** while the corresponding eigenvalues occupy the diagonal elements of the matrix **pcaLatentRC**.  In this formulation, **pcaScoreRC * pcaLatentRC * pcaCoeffRC'** reconstructs the matrix **rescaledRealizationRC.**  As a result, synthetic (simulated) data vectors $\vec{d}$ can be generated by sampling the $m$:th element, $d_m$, from a Gaussian distribution with variance $\lambda_m$, where $\lambda_m$ is the $m$:th diagonal element of **pcaLatentRC**.  The synthetic data vector is then obtained by right-multiplying each $\vec{d}$ by **pcaLatentRC * pcaCoeffRC'**, adding back **realizationSaveMeanR** and then, in the case of a discrete spin system with spin values ±1, replacing positive and negative values by +1 and -1, respectively.  This procedure is however greatly simplified by noting that the elements of the $m$:th principal component vector, e.g. the $m$:th column of **pcaScoreRC** possesses the variance $\lambda_m$ by construction. Therefore, the $d_m$ can be obtained by randomly sampling the elements of this vector. The code lines for this procedure are

```
for sampleLoop = 1 : numberOfSimulatedSamples
   indexVectorR = randi( numberOfRealizations, 1, numberOfPoints );
   for weightsLoop = 1 : numberOfPoints
      syntheticWeightsR(weightsLoop) = u(indexVectorR(weightsLoop), weightsLoop);
   end
   syntheticDataRC(sampleLoop, :) = syntheticWeightsR * pcaLatentRC * pcaCoeffRC' + ...
         realizationSaveMeanR;
end
syntheticDataRC( syntheticDataRC > 0 ) = 1;
syntheticDataRC( syntheticDataRC <= 0 ) = -1;
```

It should be noted that the quantities **coeffRC**, **scoreRC**, and **latentRC** that are returned by the **pca( )** routine are defined somewhat differently than in **svd( )** so that **scoreRC * coeffRC'** must be substituted for **pcaScoreRC * pcaLatentRC * pcaCoeffRC'** above.

**Computational Results:** The first of the PCA applications identifies physical symmetries in recorded data. Following the strategy introduced in [39], a total of 500 random two-dimensional harmonic oscillator trajectories with $k = m = 1$ were generated and propagated to $t = 4000$ with initial radii evenly distributed between $r = 2$ and $r = 3$ and initial velocities lying in a square with corner positions $\vec{v} = (1,1)$ and $\vec{v} = (-1,-1)$. (The initial positions must be sufficiently distant from the coordinate origin that a strong correlation is not introduced between position and angular momentum, decreasing the ratio between the smallest and largest PCA variances.) The position and velocity variables were recorded at each algorithmically generated interval, resulting in $4 \times 10^6$ data points. The physically conserved quantities are then encoded in the principal component vectors with the smallest variances as these coincide with directions in variable space along which the result is effectively invariant. In the present case the conserved quantities can immediately be identified from the first column of Table 1 (the second highest order principal component vector) as the fifth minus the sixth variables from the next to last principal component vector, corresponding to $xv_y - yv_x$ and from the last column in the table as the sum of the first four variables yielding $x^2 + y^2 + v_x^2 + v_y^2$. The variances of all principal components are presented for completeness in Table 2. Note that this method is somewhat analogous to polynomial regression, which was employed to find conserved quantities in [39]. Therefore, while it can identify symmetries that can be expressed algebraically as low-order polynomials, non-polynomial conserved quantities again cannot be easily identified.

| | |
|---|---|
| 0.0201 | 0.5068 |
| 0.0447 | 0.4909 |
| 0.0201 | 0.5069 |
| 0.0447 | 0.4909 |
| -0.7053 | 0.0455 |
| 0.7053 | -0.0455 |
| 0.0114 | -0.0036 |
| 0 | 0 |
| 0 | 0 |
| 0.0114 | -0.0036 |
| 0 | 0 |
| 0 | 0 |
| 0 | 0 |
| 0 | 0 |

*Table 1: The two highest order principal components e.g. with smallest variances (eigenvalues).*

| |
|---|
| 4.3393 |
| 3.9976 |
| 2.8076 |
| 2.2801 |
| 2.2105 |
| 1.6845 |
| 1.6839 |
| 1.5548 |
| 1.5009 |
| 1.438 |
| 1.4375 |
| 1.3808 |
| 0.8087 |
| 0.4867 |

*Table 2: The variances (eigenvalues) of all principal components*

Next, the classical two-dimensional ferromagnetic Ising model on a periodic lattice of $8 \times 8$ spins with unit spins described by the Hamiltonian

$$H = J \sum_{i,j \in \text{nearest neigbhors}} S_i S_j$$

is considered. In an infinite system the $Z_2$ inversion symmetry is broken below $T = 2.269$ (temperature and energy are here expressed in normalized units $T/k_b$ and $E/J$) as the system transitions from a disordered to an ordered phase. To apply the PCA to the Ising model 12000 spin configurations are generated at a temperature of $T = 3.526$ with the Wolff cluster reversal procedure [43] which results in the joint distribution of samples of Figure 1, in which energy and magnetization are displayed on the horizontal and vertical axes, respectively. The specific heat per unit spin, calculated from the energy fluctuations among the different samples according to

$$c = -\frac{<E>^2 - <E^2>}{T^2}$$

equals 0.2537 for the samples employed, which compares to the exact result, $c = 0.2556$. [1]

Subsequently, each spin configurations is flattened into a vector with elements that are 1 or -1 corresponding to up and down spins, respectively and these vectors are stacked in successive rows of the $12000 \times 64$ matrix $\mathbf{M}$. The principal components and their respective eigenvalues (variances) are the solutions to the eigenproblem $\mathbf{M^T M} \vec{\chi}_i = \lambda_i \vec{\chi}_i$. Since this coincides with the matrix equation encountered in the least squares procedure, performing the principal value decomposition and discarding the highest order eigenvectors and eigenvalues yields an optimal least squares approximation to the data for the given number of retained principal components. [18] Each data sample, $\vec{d}$ is represented in the space of principal components by the projection, $\mathbf{M}\vec{d}$. This procedure can be contrasted with the low dimensional representation of the Ising mode generated in [13] through stochastic neighbor embedding.

The structure of the principal values possesses several intriguing features which are somewhat analogous to those observed for variational autoencoders in [25]. In particular, if the magnitudes of the principal values are plotted against their index, the resulting curve, Figure 2, displays a pattern of groups of four closely spaced principal values. This would appear to be associated with the geometry of the boundary and in particular the distortion of the square lattice when projected onto a sphere, which encodes the lack of rotational symmetry. That is, while the lattice is periodic, the distance from the center point back to itself is a minimum along the direction of the coordinate axes and a maximum along the lattice diagonals, providing an implicit rectangular structure. To illustrate, Figure 3, displays the magnitudes of the principal values for a $8 \times 4$ periodic Ising lattice, which are seen to group into sets of 2 and 4, consistent with the coincidence of the width with twice the length.

Considering next the evolution of the principal components with temperature, Figure 4 displays the ratio, $R$, of the logarithm of the lowest to the second lowest order principal component as the temperature is increased from 1.2 to 4 in steps of 0.1 in normalized units. This ratio is large and linearly varying at low temperatures, where all the spins are nearly aligned and therefore can be characterized by a single principal component, while it is small and approximately constant for low temperatures. The transition between these two regions occurs near the critical temperature. By trial and error, it was found that in this particular example

$$-T^2 \frac{d \log R}{dT}$$

possesses a maximum close to the critical temperature, as demonstrated by Figure 5.

As outlined in the previous section, if the singular value decomposition of the difference of the original data matrix with a vector $\vec{m}$ containing the mean of its columns is written as $\mathbf{suv^T}$, a synthetic data vector can be generated in a two-step process. First, an input vector, $\vec{d}_{\text{random projection}}$, is constructed by randomly selecting a value from each of the columns of $\mathbf{s}$ (as $\mathbf{s}$ is composed of the projections of the input data onto the principal vectors). The synthetic data sample is then obtained by inverting the PCA according to

$$\vec{d}_{\text{synthetic}} = \vec{d}_{\text{random projection}} \mathbf{uv^T} + \vec{m}$$

where $\mathbf{u}$ only retains the $N_{\text{principal}}$ lowest order principal values (eigenvalues). The probability distribution functions of the synthetic data generated in this manner are displayed in Figure 6 and Figure 7 for $N_{\text{principal}} = 20$ and $N_{\text{principal}} = 64$ respectively. While both of these appear similar to that of the input data, e.g. Figure 1, the result for $N_{\text{principal}} = 20$ is displaced to smaller energies. Moreover, the specific heat obtained from the synthetic data equals 0.5745 for $N_{\text{principal}} = 20$ and falls monotonically to 0.339 for $N_{\text{principal}} = 64$. Evidently, the long range spin correlations near the critical temperature lead to correlations among the projections of the states onto the different principal axes, which are however neglected by this technique.

**Conclusions:** The properties of principal component analysis in the context of spin systems presented in this paper suggest several topics for further investigation. For example, the ability of the PCA to identify symmetries in physical systems provides a simple alternative to previously suggested machine learning techniques. However none of these procedures have been applied to symmetries that cannot be expressed in a simple algebraic form. While such cases could perhaps be handled by a judicious choice of basis functions or coordinate transformations, automating such procedures appears inherently challenging. A second issue is the determination of conserved quantities when additional correlations between the variables are introduced through the initial conditions. While each symmetry will then not be encoded in a separate principal component vector, an algorithm can likely be found for extracting the symmetries from combinations of these vectors.

Concerning the PCA analysis of the Ising model, the apparent influence of the boundary geometry on the principal values might perhaps find application to inverse problems as global features of the boundaries could be partly related to the principal component eigenvalues. Additionally, much along the lines of the discussion in [39], the behavior of the ratio of the lowest and second order principal components clearly signals the presence of a phase transition. However, considerable analysis would presumably be required to quantify the exact dependence, if one exists. Finally, while the synthetic data generated by inverting the principal component decomposition adequately regenerates the joint energy-magnetization probability density distribution of the input data, the associated specific heat values display large errors. This discrepancy reflects correlations between different projections of the data vectors along the principal component axes that are neglected when these values are manipulated as statistically independent quantities. A detailed analysis of the transformation of the correlations among the projections onto the principal axes and a numerical method for retaining these correlations in the synthetic data generation algorithm would accordingly be of considerable theoretical and practical interest.

**Acknowledgements:** The Natural Sciences and Engineering Research Council of Canada (NSERC) is acknowledged for financial support.

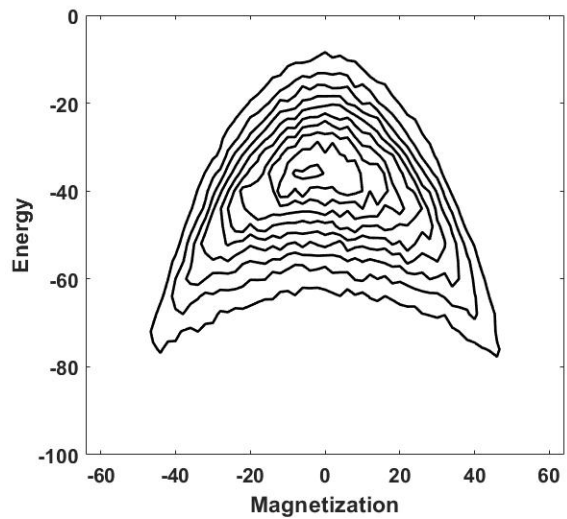

*Figure 1: The joint energy-magnetization distribution of the input $8 \times 8$ Ising model data at $T = 3.526$*

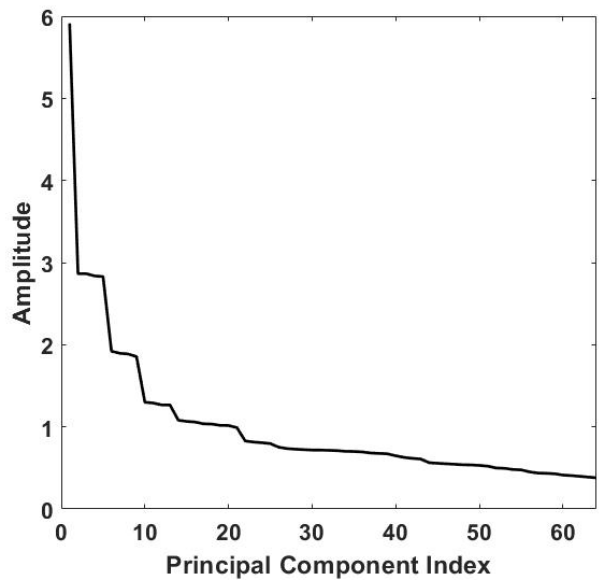

*Figure 2: The amplitude of the principal components for the $8 \times 8$ Ising model at T=3.526*

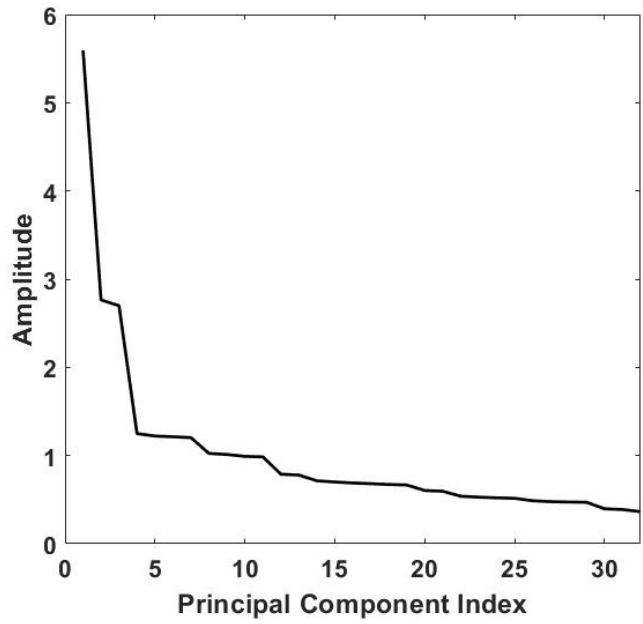

*Figure 3: As in Fig. 3, but for a 8 × 4 spin Ising model*

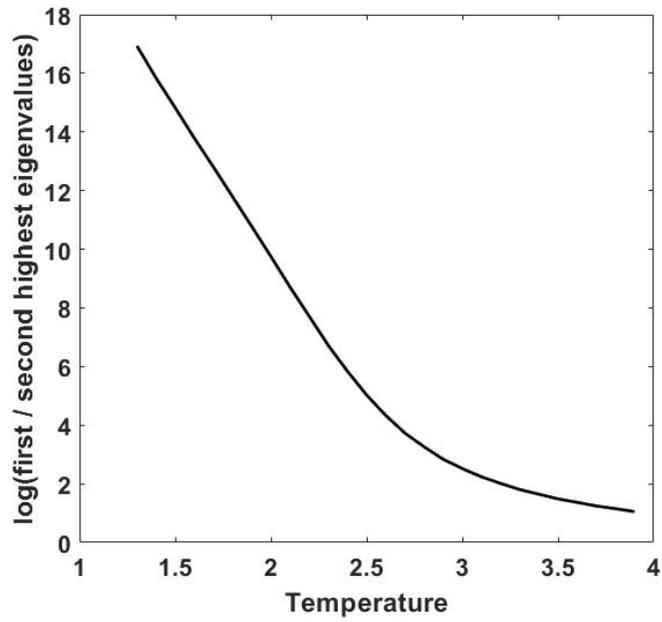

*Figure 4: The logarithm of the ratio of the lowest to second lowest principal component amplitude for the input data of Fig. 1*

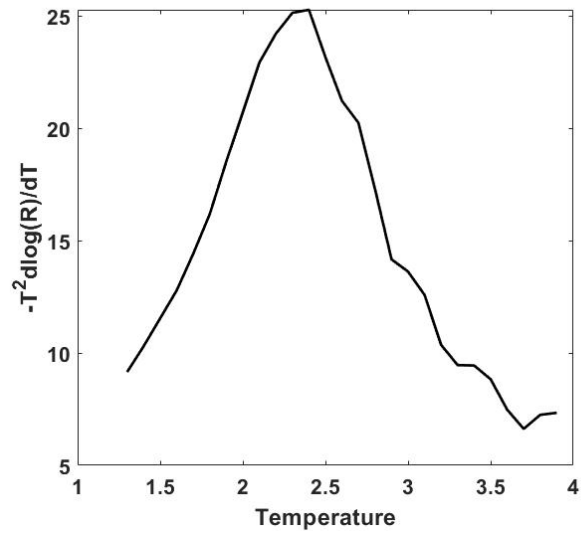

*Figure 5: The derivative of the curve of the previous figure multiplied by $-T^2$.*

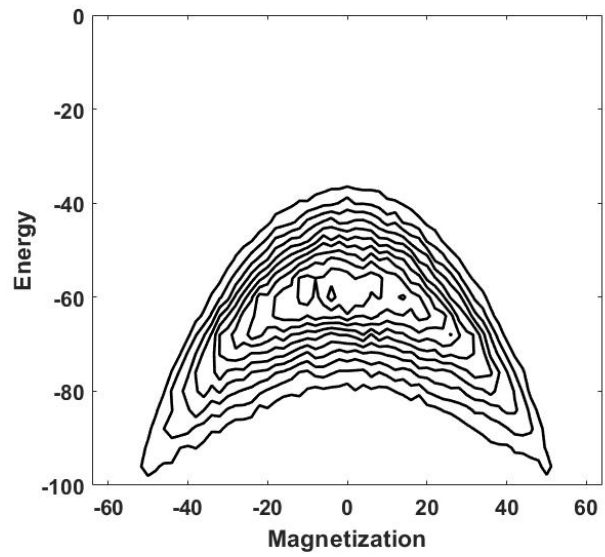

*Figure 6: The pdf of the synthetic data when 20 principal components are retained.*

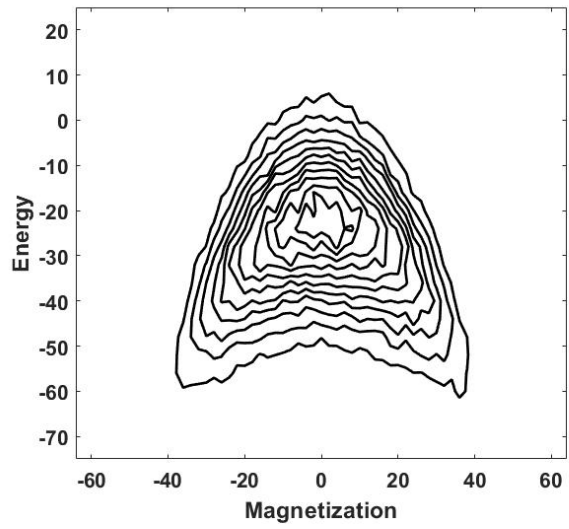

*Figure 7: As in Fig. 6 but employing all 64 principal components*